\documentclass[pra,showpacs,twocolumn]{revtex4}
\usepackage{bm,graphicx,amsmath}
\begin{document}
\title{Spin-orbit coupled Bose-Einstein condensate in a tilted
optical lattice}
\author{Jonas Larson}
\email{jolarson@fysik.su.se}
\affiliation{NORDITA, Se-106 91
Stockholm, Sweden}
\affiliation{Department of Physics, Stockholm University,
AlbaNova University Center, SE-106 91 Stockholm, Sweden}
\author{Jani-Petri Martikainen}
\affiliation{NORDITA, Se-106 91
Stockholm, Sweden}
\author{Anssi Collin}
\affiliation{NORDITA, Se-106 91
Stockholm, Sweden}
\author{Erik Sj\"oqvist}
\affiliation{Department of Quantum Chemistry,
Uppsala University, Box 518, Se-751 20 Uppsala, Sweden}
\affiliation{Centre for Quantum Technologies, National University
of Singapore, 3 Science Drive 2, 117543 Singapore, Singapore}

\date{\today}

\begin{abstract}
Bloch oscillations appear for a particle in a weakly tilted periodic
potential. The intrinsic spin Hall effect is an outcome of
a spin-orbit coupling. We demonstrate that both these phenomena can be
realized simultaneously in a gas of weakly interacting ultracold atoms
exposed to a tilted optical lattice and to a set of spatially dependent
light fields inducing an effective spin-orbit coupling. It is found that
both the spin Hall as well as the Bloch oscillation effects may coexist,
showing, however, a strong correlation between the two. These correlations
are manifested as a transverse spin current oscillating in-phase with the
Bloch oscillations. On top of the oscillations originating from the periodicity
of the model, a trembling motion is found which is believed to be atomic
{\it Zitterbewegung}. It is argued that damping of these Zitterbewegung
oscillations may to a large extent be prevented in the present setup considering
a periodic optical lattice potential.
\end{abstract}
\pacs{03.75.Mn, 67.85.Hj, 71.70.Ej}
\maketitle

\section{Introduction}
Recent years have seen an increased interest in many-particle spin
dynamics, also termed {\it spintronics}~\cite{spintronics}. Most of
this research have been devoted to condensed matter systems, where the
electron spin constitutes the spin degree of freedom~\cite{solidspintronics}.
However, related or equivalent phenomena may occur by considering the
polarization of light~\cite{lightspintronic} or internal Zeeman levels
of cold atoms exposed to optical light fields~\cite{atomspintronic} or
spatially varying magnetic fields~\cite{magn}.

In many of the spintronics schemes developed in the condensed matter community,
a spin-orbit (SO) coupling plays a crucial role for the dynamics. The particular
form of the SO coupling can be of either Rashba~\cite{rashba} or
Dresselhaus~\cite{dres} type, both being frequently analyzed in terms
of an effective gauge force~\cite{spin1,anhall1}. In other words, one can
interpret the SO couplings as giving rise to an effective spin-dependent
Lorentz-type force. Thus, the SO coupling can serve as a spin-filter or a
Stern-Gerlach apparatus. As the manufacturing techniques improves, undesired 
effects deriving from crystal impurities can be greatly suppressed and
several successful experiments demonstrating the SO coupling and its outings 
have been reported recently~\cite{anhallrev}. Ultracold atoms in optical 
lattices~\cite{maciek,bloch08} possess another possibility to implement 
effective SO couplings, and the study of it within a clean environment. While 
in condensed matter settings where the coupling is an intrinsic property of 
the systems, for cold atoms it has to be externally created via spatially 
varying laser fields.

Ever since the pioneering experiments on Bose-Einstein condensation
(BEC)~\cite{bec}, the field has witnessed a tremendous progress both experimentally
and theoretically~\cite{bec2}. The dispersive dipole coupling between light and
atoms enables the possibilities to study matter waves in periodic potentials
and the experimental realization of model systems discussed within theoretical
solid state physics~\cite{maciek}. Today, both the weakly interacting atomic
gases as well as more strongly interacting ones can be realized experimentally.
Furthermore, utilizing the internal atomic hyperfine levels enable studies of
the interplay between internal and external degrees of freedom~\cite{spinor}.
One interesting feature is the realization of artificial gauge potentials by
means of coupling of internal atomic Zeeman levels~\cite{lightgauge,ohberg,lightexp1,lightexp2}.
The technique of light induced gauge potentials allows for non-Abelian behavior to
be analyzed and the SO couplings are one such example~\cite{galitsky,jonaserik}.

In this paper, we investigate the dynamics of a spinor BEC subjected
both to a two-dimensional weakly tilted optical square lattice as well to a
set of spatially varying light fields giving rise to an effective SO coupling
between the internal atomic states. Thereby, its dynamics will be driven by
two mechanisms, namely; (i) Bloch oscillations (BOs) arising due to the
tilting of the lattice, and (ii) an intrinsic spin Hall effect (SHE) deriving
from the effective SO coupling. The two phenomena may be analyzed in terms
of an extrinsic force along the longitudinal direction of the lattice tilt and an
intrinsic transverse force. Being orthogonal to each other, these forces
render a complex correlated motion of the BEC in the lattice potential. Furthermore,
the particular form of the SO coupling renders Dirac cones characterized by linear
dispersions reminiscent of relativistic electrons. As an outcome of these, the
evolution possesses as well a trembling {\it Zitterbewegung} motion~\cite{zitt}.
Normally, such effect damps out very fast \cite{solano}, but we show that in a
lattice situations as in the present paper it may indeed be long-lived and
hence more easily observable.
Moreover, the optical lattice can serve as trapping potential for the atoms in the BO regime
and no confining Harmonic trap is needed which would typically prevent buildup of spin currents~\cite{trap}.

While SO couplings arise as an intrinsic property in many crystals, in cold atom 
systems it has to be implemented externally. This, at the one hand may be a source 
for experimental errors, but at the other it gives a handle for the SO coupling; its 
strength and form can be tuned experimentally. The same holds for the optical lattice, 
the lattice strength and geometry are both control parameters, which has made it 
possible to observe BOs for BECs in optical lattices~\cite{blochexp2}. Due to the 
large atomic mass compared to the mass of the electron, Zitterbewegung is expectedly 
more clearly manifested in atomic systems. The first experiments detecting Zitterbewegung 
in atomic systems were recently carried out~\cite{solano,vaishnav08}. A different but 
related setup can be found in Ref.~\cite{magn}, where a spinor condensate accelerating 
in a periodic magnetic field was studied. We note that in Ref.~\cite{magn}, the 
spatially dependent magnetic field gives rise to an effective gauge force. However, Bloch 
oscillations were not considered, and moreover, in the present work the periodic potential is 
external and independent of the SO coupling.

The outline of the paper is as follows. In Sec.~\ref{sec2}, we introduce the model
system and outline how effective SO couplings are rendered in such systems. To
start with we analyze the free case in which the problem is analytically solvable
and demonstrate the SHE and the Zitterbewegung effect. Applying the acceleration
theorem we also solve the problem under constant acceleration. The dispersions,
once an optical lattice has been included, are solved numerically displaying the
Dirac cones. Section~\ref{sec3} presents the main analysis. The full numerical
results are presented, where we demonstrate the presence of a SHE and how it is
modified by the longitudinal oscillatory motion. In particular, it is shown that
the Hall current will also oscillate, however with an increasing oscillation
amplitude. On top of the oscillations deriving from the motion in a periodic
potential, we find a persisting Zitterbewegung effect. Finally, we conclude
with a summary in Sec.~\ref{sec4}.

\section{Model system}\label{sec2}
Rashba~\cite{rashba} and Dresselhaus~\cite{dres} SO couplings
appear in a variety of solids, ranging from
semiconductors~\cite{semiSO}, thin magnetic films~\cite{magSO}, to
graphene~\cite{grapSO}. The SO couplings bring about a degeneracy in
the energy dispersions. Around these degeneracies, the dispersions
are linear, reminiscent to that of relativistic particles,
and therefore they are called {\it Dirac cones}. The relativistic
properties have been confirmed experimentally in graphene either by 
{\it angle resolved photoemission spectroscopy} or by
considering the conductivity and cyclotron motion of electrons~\cite{rel0,rel1}. The SO
coupling and its corresponding degeneracy induces as well a non-trivial
effective gauge structure, giving rise to phenomena such as the intrinsic
spin and anomalous Hall effects~\cite{spin1,anhall1}. In the field of cold 
atoms, the first light induced effective magnetic fields were demonstrated 
in Ref.~\cite{lightexp2}, and signatures of non-Abelian structures were recently 
observed, see~\cite{nonab} and reference therein. BOs with ultracold atoms have 
already been realized experimentally by numerous groups~\cite{blochexp2,blochexp1}.
We proceed by presenting a scheme which realizes the desired SO coupling in
an atomic tripod system.

\subsection{Light induced effective spin-orbit coupling}\label{ssec3a}
The idea of effective gauge potentials arising from adiabatic evolution dates
back to the seminal works by Mead on vibrations of molecules~\cite{mead}.
Berry~\cite{berry} generalized this finding by demonstrating the existence
of an effective gauge structure in the context a quantum adiabatic evolution.
Wilczek and Zee \cite{wz} further generalized Berry's result and found an effective
non-Abelian gauge potential accompanying adiabatic evolution of degenerate
energy eigenstates. Dum and Olshanii \cite{lightgauge} showed that the effective
gauge potentials can be generated in systems of ultracold
atoms subjected to spatially dependent light fields that couple
internal atomic states. Following an approach for creating
non-Abelian gauge potentials in time-dependent systems~\cite{bergmann},
Ruseckas {\it et al.}~\cite{ohberg} outlined a scheme how to achieve
effective non-Abelian gauge potentials by considering an atomic
tripod system in spatially dependent light fields. The coupled four-level
system possesses two
degenerate dark states that, within certain parameter regimes, can be
decoupled from the remaining two bright states via standard
adiabatic elimination. The flexibility of this tripod
system, deriving from different laser configurations, allows for a
variety of phenomena to be studied, such as, e.g., relativistic properties
\cite{vaishnav08,juziliunas08,hou09}, SHEs \cite{app2}, novel phases of
the BEC \cite{app3}, and topological defects as effective magnetic monopoles 
\cite{app4} or vortices~\cite{nonab}.

\begin{figure}[ht]
\begin{center}
\includegraphics[width=8cm]{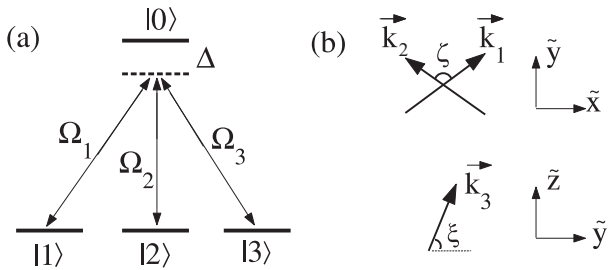}
\caption{Atomic (a) and laser (b) configurations. The specific laser arrangement,
made up of standing and traveling waves, is described in detail in the text.}
\label{fig4}
\end{center}
\end{figure}

In the present paper we consider a tripod setup similar to the one
described in Refs.~\cite{galitsky,jonaserik}. The details of
atomic and laser configurations rendering the gauge potentials can be
seen in Fig.~\ref{fig4}. The internal dynamics, is driven by three lasers
that couple the three lower atomic states $|1\rangle$, $|2\rangle$, and $|3\rangle$
to the excited state $|0\rangle$. Parameterizing the laser fields as
$\Omega_1=\Omega\sin\theta\cos\varphi e^{iS1}$,
$\Omega_2=\Omega\sin\theta\sin\varphi e^{iS_2}$, and
$\Omega_3=\Omega\cos\theta e^{iS_3}$, such that
$\Omega=\sqrt{\left|\Omega_1\right|^2 + \left|\Omega_2\right|^2 +
\left|\Omega_3\right|^2}$, the two degenerate
dark states of the corresponding Hamiltonian can be written as
\begin{equation}
\begin{array}{lll}
|u_1\rangle & = &
\displaystyle{\frac{1}{\sqrt{2}}}\Big(\left[ e^{-i\pi/4} \cos \! \theta
\cos \! \varphi + e^{i\pi/4} \sin\varphi \right] e^{-iS_{13}} |1\rangle \\ \\
 & & + \left[ e^{-i\pi/4} \cos \! \theta \sin \! \varphi -
e^{i\pi/4} \cos \varphi \right] e^{-iS_{23}} |2\rangle \\ \\
 & & - e^{-i\pi/4} \sin\!\theta |3\rangle \Big), \\ \\
|u_2\rangle & = &
\displaystyle{\frac{1}{\sqrt{2}}} \Big( \left[ e^{-i\pi/4} \cos \! \theta
\cos \! \varphi - e^{i\pi/4} \sin\varphi \right] e^{-iS_{13}} |1\rangle \\ \\
 & & +\left[ e^{-i\pi/4} \cos \! \theta \sin \! \varphi + e^{i\pi/4} \cos\varphi
\right] e^{-iS_{23}} |2\rangle \\ \\
 & & - e^{-i\pi/4} \sin \! \theta |3\rangle \Big), \\ \\
\end{array}
\end{equation}
where $S_{kl}=S_k-S_l$. Provided the field amplitude
$\Omega$ is much larger than the atom-light detuning $\Delta$ (as
depicted in Fig.~\ref{fig4}), the two dark states can be safely eliminated from the two
bright states, resulting in an effective
two-level system~\cite{ohberg}. To derive an effective SO coupling between
the two internal dark states $|u_1\rangle$ and $|u_2\rangle$, we choose
$S_1=S_2$, $S_{31}=v_sy$, $\varphi=v_\varphi x$, and $\theta\in[0,\pi]$.
This can be achieved by coupling the two ground states $|1\rangle$
and $|2\rangle$ to the excited state $|0\rangle$ with two lasers intersecting
with an angle $\zeta$ and propagating in the $xy$-plane. These lasers correspond
to a standing wave in the $x$-direction and a plane wave in the $y$-direction.
A third laser, couples the states $|3\rangle$ and $|0\rangle$, and is a plane
wave in the $yz$-plane with an angle $\xi$ of propagation relative to the
$y$-axis. In terms of the angles $\zeta$ and $\xi$, and the wave
numbers $k_j = |\vec{k}_j|$, $j=1,2,3$, we have
\begin{equation}
\begin{array}{l}
S_1=S_2=k\tilde{y}\cos (\zeta/2), \\ \\
S_3=k_3\tilde{y}\cos\xi , \\ \\
\varphi=2k\tilde{x}\sin (\zeta/2),
\end{array}
\end{equation}
with $k\equiv k_1=k_2$. Performing the adiabatic elimination, taking into account that
$\Omega\gg|\Delta|$ and the fact that we consider ultracold atoms so
that excitations due to atomic motion can be neglected, the effective
single particle two-level Hamiltonian becomes in dimensionless variables \cite{galitsky}
\begin{equation}\label{ham4}
\hat{H}_{\mathrm{eff}}^{(0)}\!=\! \hat{p}_x^2+
\hat{p}_y^2+\hat{p}_z^2
+\delta_0\hat{\sigma}_z-v_0\hat{\sigma}_x \hat{p}_x -
v_1\hat{\sigma}_y \hat{p}_y,
\end{equation}
where $\delta_0$ is an effective Zeeman splitting,
$v_0=(v_\varphi/2)\cos\theta$, $v_1=(v_s/2)\sin^2\theta/2$. The Pauli matrices
$\hat{\sigma}_j$, $j=x,y,z$, operate in the space spanned by the dark
states, i.e., $\hat{\sigma}_x =|u_1\rangle \langle u_2| + |u_2\rangle \langle u_1|$,
$\hat{\sigma}_y = -i|u_1\rangle \langle u_2| + i|u_2\rangle \langle
u_1|$, and $\hat{\sigma}_z = |u_1\rangle \langle u_1| - |u_2\rangle
\langle u_2|$. The $p$-dependence of the last two terms originates from non-adiabatic
couplings~\cite{jonas1}. We note that these do not include $p_z$ and as a consequence
the $z$-dependence can be fully factored out. In Eq.~(\ref{ham4}), we have used
dimensionless parameters. Explicitly, when we further on introduce a lattice potential
whose wave number is $k$, we let the characteristic length scale be $l=2\pi k^{-1}$ and
we let the characteristic energy scale be the recoil energy $E_R=\hbar^2k^2/2m$. Consequently,
time $t$ is measured in units of $\tau=\hbar/E_R$. The superscript $(0)$ indicates the free
Hamiltonian in absence of a spatially dependent optical lattice potentials.

The Zeeman shift $\delta_0$ splits the degeneracy of the two internal states and together
with the SO coupling it gives rise to a transverse current, an intrinsic anomalous Hall
effect. We will, however, focus on the intrinsic SHE obtained when $\delta_0=0$. In this
case, the transverse particle current vanishes but, on the other hand, a transverse spin
current persists. The $\delta_0=0$ condition can be achieved by internal Stark shift of
the individual bare atomic levels~\cite{galitsky,ohberg2}. Thus, in
the following we will assume $\delta_0=0$, and furthermore we will also
consider the symmetric SO coupling characterized by $v_0=v_1\equiv v_{so}$.
It should be pointed out, however, that even by lifting these constrains
an interplay between transverse and longitudinal atomic motion will still occur.

One important prerequisite for successful application of the current scheme
is that the system can be prepared in the dark subspace spanned by
$\{|u_1\rangle,|u_2\rangle\}$. This issue has been discussed
in~\cite{ohberg3}. The idea is to slowly turn on the
external lasers $\Omega_k$, $k=1,2,3$, such that the state
follows adiabatically states in the dark subspace, preparing
a superposition of $\{|u_1\rangle,|u_2\rangle\}$ but no
bright state components.

\subsection{Analytical results without a lattice potential}

We note that we may rewrite the Hamiltonian in Eq.~(\ref{ham4}) ($\delta_0=0$) as
\begin{equation}\label{ham2}
\hat{H}_{\mathrm{eff}}^{(0)}=(\hat{p}_x-\hat{\mathcal{A}}_x)^2+(\hat{p}_y -
\hat{\mathcal{A}}_y)^2+\hat{\Phi}
\end{equation}
with the vector potential
\begin{eqnarray}
(\hat{\mathcal{A}}_x,\hat{\mathcal{A}}_y) =
-v_{so}(\hat{\sigma}_x,\hat{\sigma}_y)/2
\label{eq:vp}
\end{eqnarray}
and the scalar potential
\begin{eqnarray}
\hat{\Phi} = -\frac{1}{2} v_{so}^2 \hat{1} .
\end{eqnarray}
These potentials constitute vector and scalar parts of the effective
gauge potential. Although being constant in space, the vector potential
gives rise to a non-trivial effective gauge field due to its non-Abelian
structure, $[\hat{\mathcal{A}}_x,\hat{\mathcal{A}}_y]\neq0$. Explicitly,
this effective field is given by the antisymmetric field tensor
\begin{equation}
\hat{\mathcal{F}}_{kl} = \partial_k\hat{\mathcal{A}}_l-\partial_l
\hat{\mathcal{A}}_k-i[\hat{\mathcal{A}}_k,\hat{\mathcal{A}}_l] ,\,\, \ k,l=x,y,
\end{equation}
resulting in $\hat{\mathcal{F}}_{xy} = \frac{1}{2} v_{so}^2 \hat{\sigma}_z$ and
$\hat{\mathcal{F}}_{kl} = 0$ otherwise. As a result, the spin-orbital
motion of the electrons can be understood in terms of an effective Lorentz-type
force, akin to that found by Wong~\cite{wong} for a color-charged
classical particle moving in a non-Abelian gauge field. The corresponding force
operator may be derived from Heisenberg equations of motion according
to~\cite{wpspinhall}
\begin{equation}\label{lforce}
\begin{array}{lll}
\hat{\bf{F}}_{SO} & = &
\displaystyle{\frac{1}{2}\frac{d^2\hat{\bf{r}}}{dt^2} =
\frac{1}{2}[\hat{H}_0,[\hat{\bf{r}},\hat{H}_0]]}\\ \\
 & = &
\displaystyle{v_{so}^2 (\hat{\bf{p}} \times {\bf e}_z) \hat{\sigma}_z},
\end{array}
\end{equation}
where it is understood that the operators on the right-hand side are all given in
the Heisenberg representation and the prefactor $\frac{1}{2}$ in front of the
acceleration is the mass in the chosen units. Note that the force in the
$y$-direction is proportional to $-\hat{p}_x$ and the force in the $x$-direction
is proportional to $\hat{p}_y$. This demonstrates the transverse character of
$\hat{\bf{F}}_{SO}$ and its equivalence to a Lorentz or Coriolis force. The
crucial aspect of Eq.~(\ref{lforce}) is, however, that the force is state
dependent: atoms in dark state $|u_1\rangle$ feel a force with
opposite sign compared to those in dark state $|u_2\rangle$. Therefore,
the motion of the particles in the $xy$-plane may separate
the two spin states implying that a non-zero spin current is generated.
In the following, we shall call the $x$-direction
longitudinal and the $y$-direction transverse.

The effect of the above Lorentz-type gauge force can be examined by
computing how the initial ($t=0$) state
\begin{equation}
\Psi(p_x,p_y,0)\!=\!
\psi_1(p_x,p_y,0) |u_1\rangle + \psi_2(p_x,p_y,0)
|u_2\rangle
\end{equation}
evolves in time under influence of $\hat{H}_{\mathrm{eff}}^{(0)}$. We assume
that the internal and momentum states of the atom are initially uncorrelated
and that the initial momentum states are Gaussian wave packets of the form
\begin{eqnarray}
\label{gaussin}
\psi_{1}(p_x,p_y,0) & = & - \psi_{2}(p_x,p_y,0) = \psi(p_x,p_y)
\nonumber \\
 & = & \frac{1}{\sqrt[4]{4\pi\Delta_p^2}} e^{-\frac{(p_x-p_0)^2+p_y^2}{2\Delta_p^2}},
\end{eqnarray}
where $\Delta_p$ sets the momentum uncertainty and $p_0$ is the average
momentum along the $x$-direction. This choice corresponds to the pure
internal atomic state $|-\rangle = (|u_1\rangle-|u_2\rangle)/\sqrt{2}$.
Given this initial state, the time evolved state is readily obtained as
\begin{eqnarray}
\Psi(p_x,p_y,t) & = &
\psi_1(p_x,p_y,t) |u_1\rangle +
\psi_2(p_x,p_y,t) |u_2\rangle
\end{eqnarray}
with
\begin{eqnarray}\label{timesol}
\psi_1(p_x,p_y,t) & = &
\frac{1}{\sqrt{2}}\left[\cos(\lambda t) + i\sin(\lambda t)e^{-i\theta}\right]
\nonumber \\
 & & \times \psi(p_x,p_y) e^{-i(p_x^2+p_y^2)t} ,
\nonumber \\
\psi_2(p_x,p_y,t) & = &
-\frac{1}{\sqrt{2}}\left[ \cos(\lambda t) + i\sin(\lambda t)e^{i\theta}\right]
\nonumber \\
 & & \times \psi(p_x,p_y) e^{-i(p_x^2+p_y^2)t} .
\end{eqnarray}
Here, we have used the energy dispersions $E_{\pm}$ and the energy eigenstates
$\phi_{\pm}$ of the Hamiltonian in Eq.~(\ref{ham4}), which read
\begin{eqnarray}
E_\pm(p_x,p_y) & = & p_x^2 + p_y^2 \pm \lambda ,
\nonumber \\
|\phi_{\pm} \rangle & = & \displaystyle{\frac{1}{\sqrt{2}}
\left( \pm |\uparrow\rangle + e^{i\theta} |\downarrow\rangle \right)},
\label{eig}
\end{eqnarray}
where $\lambda = v_{so} \sqrt{p_x^2+p_y^2}$ and $\tan(\theta)=p_y/p_x$.

\begin{figure}[ht]
\begin{center}
\includegraphics[width=8cm]{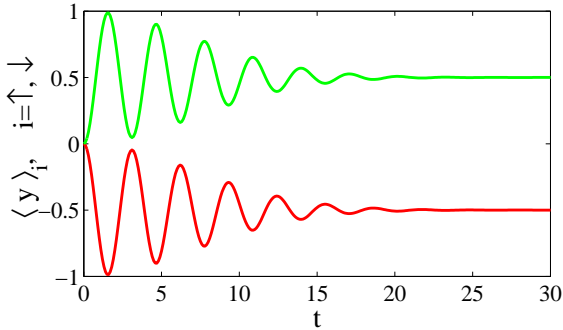}
\caption{(Color online) Time evolution of
$\langle\hat{y}\rangle_{1,2}$ (green curve correspond to
the dark state $|u_1\rangle$ and the red curve correspond to the dark
state $|u_2\rangle$) as defined in Eq.~(\ref{analyexp}). The
dimensionless parameters in Eqs.~(\ref{ham4}) and (\ref{gaussin}) are
$v_{so}=1$ ($=v_1=v_2$), $\delta_0=0$, $p_0=1$, and $\Delta_p=0.1$. The
initial state is thus a Gaussian with a non-zero average momentum along
the $x$-direction.}
\label{fig1}
\end{center}
\end{figure}

In Fig.~\ref{fig1} we display the dimensionless averages
\begin{eqnarray}\label{analyexp}
\langle\hat{y}\rangle_i =
2i \iint \psi_i^{\ast} (p_x,p_y,t) \frac{\partial}{\partial p_y}
\psi_i (p_x,p_y,t) dp_x dp_y ,
\end{eqnarray}
with $i=1,\,2$. The factor $2$ derives from normalization of
the individual wave packet components. Since the initial velocity is along the
longitudinal $x$-direction, the deviation from $\langle \hat{y} \rangle_i=0$
is a direct consequence of the state-dependent Lorentz-type gauge force $\hat{\bf{F}}_{SO}$.
Note that $\langle\hat{y}\rangle_1 = -\langle \hat{y} \rangle_2$
giving a vanishing transverse particle current. This symmetry is broken by
keeping the Zeeman term $\delta_0\hat{\sigma}_z$ in the Hamiltonian of Eq. (\ref{ham4}).
The particle current induced by such a $\hat{\sigma}_z$-term goes under the
name intrinsic anomalous Hall effect. After $t\approx20$, the separation of
the two internal dark states has reached an asymptotic value. Until $t\approx20$,
the transverse spin current shows an oscillating behavior deriving from the same
mechanism as the one underlying relativistic Zitterbewegung in the Dirac
equation~\cite{zitt}. In the Dirac equation the trembling motion describes
coherent coupling between positive and negative energy solutions, while in
the present situation it describes population swapping between the internal
states. In solids, the amplitude of these oscillations become extremely
small due to the lightness of the electron, while in atomic or ionic systems
the heavier masses imply larger amplitudes and possible detection~\cite{solano,vaishnav08}.
Despite heavier masses, the rapid damping of the trembling motion as depicted
in Fig. 1 is a limitation also in atomic/ionic Zitterbwegung
experiments~\cite{solano,vaishnav08}. In the next subsection we will argue
how this decay may be prevented by studying the effect in the presence
of an optical lattice.

\begin{figure}[ht]
\begin{center}
\includegraphics[width=8cm]{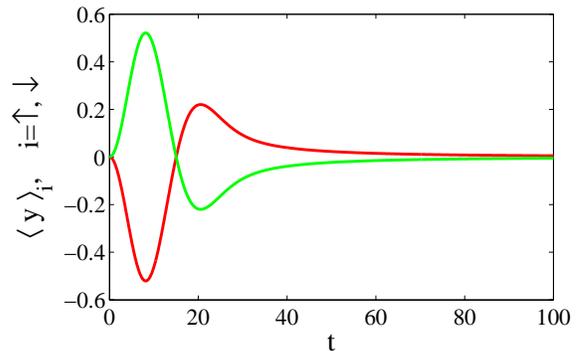}
\caption{(Color online) Same as Fig.~\ref{fig1}, but with a constant weak force
of magnitude $F_{ex}$ in the $x$-direction added to the Hamiltonian (\ref{ham4})
and assuming the validity of the adiabatic approximation. The dimensionless
parameters are $v_{so}=1$, $\delta_0=0$, $p_0=0$, $\Delta_p=0.1$, and $F_{ex}=0.01$.}
\label{fig2}
\end{center}
\end{figure}

In order to generate a current in the system one applies a constant longitudinal
force $F_{ex}\mathbf{e}_x$, e.g., induced by gravitation in atomic systems. In
the particular case of solids, the force derives from a voltage across the solid,
and the electrons reach a steady state motion in which the applied force is balanced
by scattering against crystal impurities. On the other hand, a system of cold
atoms is, to a good approximation, free from impurities and one would therefore
achieve constant acceleration. Assuming that the force is weak and that the total
time of evolution is relatively short, we can perform an adiabatic approximation,
consisting in replacing $p_x$ by $p_x-F_{ex}t$ in Eq. (\ref{timesol}). Hence,
the momentum grows linearly in time in accordance with classical motion.
How this comes about can be easily understood by considering the time
evolution of the Hamiltonian $\hat{H}=\hat{H}_0-F_{ex}\hat{x}$ for small
forces $F_{ex}$ and times $t$. In this regime, the time-evolution operator
can be approximated accordingly
\begin{equation}
\hat{U}(t)=e^{-i\hat{H}t}\approx e^{iF_{ex}\hat{x}t}e^{-i\hat{H}_0t},
\end{equation}
so that the first exponential gives the coupled time-evolution while the
second simply boosts the $x$-momentum by $-F_{ex}t$. For longer
time periods, the large velocity rendered by the acceleration will typically
cause non-adiabatic excitations, which we neglect for the moment. Similarly,
for large values of $F_{ex}t$, the factorization of the exponentials is no
longer justified. The resulting averages $\langle\hat{y}\rangle_i$ calculated
within this approximation are depicted in Fig.~\ref{fig2}. As before, we
encounter a build-up of a transverse spin current but in this case the current
changes sign after some time and finally there is no net spin separation. By
comparison with a full numerical integration of the Hamiltonian we find that
the imposed adiabatic approximation breaks down after relatively short times.
Nevertheless, numerically we have verified that the dynamics beyond this
approximation goes as well asymptotically to zero for large times. We believe
that the origin of this result is a complicated interplay between the transverse
and longitudinal forces, together with the coupling between the two internal
states. For spinless particles, the two forces render a spirally transverse
motion and hence a net transverse current. For particles with spin-dependent
Lorentz forces $\hat{\bf{F}}_{SO}$ one would expect a net
transverse spin current, but the coupled nature of the two spin states seems
to destroy such spin current in the case of constant acceleration.

\subsection{Band spectrum and the acceleration theorem}
\begin{figure}[ht]
\begin{center}
\includegraphics[width=8cm]{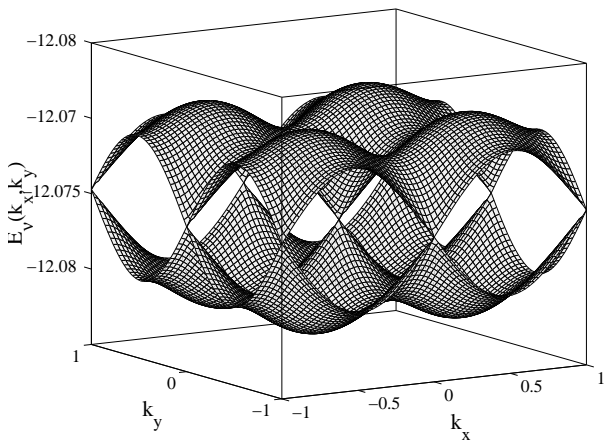}
\caption{The two lowest Bloch bands $E_\nu(k_x,k_y)$ for a square
lattice with Rashba spin-orbit coupling. In this example, $v_{so}=1$
and $V=5$. } \label{fig3}
\end{center}
\end{figure}

In the following section we numerically investigate the dynamics in the presence of
an optical lattice. Insight into the evolution behavior is gained by considering the
spectrum of the Hamiltonian including a lattice potential while disregarding the
linear force potential. Adding to the SO Hamiltonian in Eq. (\ref{ham4}) a spatially
oscillating periodic potential, yields
\begin{eqnarray}\label{ham3}
\hat{H} & = &
\hat{p}_x^2 + \hat{p}_y^2 +v_{so} \left( \hat{p}_x\hat{\sigma}_x +
\hat{p}_y\hat{\sigma}_y \right)
\nonumber \\
 & & + V\cos^2(\hat{x})+V\cos^2(\hat{y}).
\end{eqnarray}
The Hamiltonian is periodic and consequently possesses a band-gap spectrum $E_\nu (k_x,k_y)$
with associated Bloch eigenstates $u_{\nu,k_x,k_y}(x,y)$, characterized by a discrete
band-index $\nu=1,\,2,\,3,\,...$ and quasi-momentum numbers
$(k_x,k_y)$ restricted to the first Brillouin zone, i.e., $-1\leq k_x<1$
and $-1\leq k_y<1$. A detailed study of Bloch theory in the presence of SO
couplings can be found in Ref.~\cite{BlochthSO}. The first two bands
$E_1(k_x,k_y)$ and $E_2(k_x,k_y)$, obtained from numerical diagonalization
of the Hamiltonian, are shown in Fig.~\ref{fig3}. A similar band structure
has been obtained in Ref.~\cite{hou09} for cold atoms in a square optical
lattice. We directly see the effect
of the SO coupling in terms of Dirac cones at $(k_x,k_y)=(n,m)$, where
$n,\,m=,-1,\,0,\,1$. It is further seen that the ground
state is not the $u_{1,0,0} (x,y)$ Bloch state, as for a regular square
lattice lacking SO couplings. This fact is at the heart of the {\it
Jahn-Teller effect} \cite{jt}, where the Dirac cone (or {\it conical
intersection} as it is usually referred to when it appears in position
space rather than in momentum space) induces a symmetry breaking
that lowers the ground state energy. Furthermore, we note that the two
lowest bands are not gapped, while there is a gap between the second and
third band. Indeed, due to the SO coupled two-level structure of the Hamiltonian,
the bands come in pairs where each pair possesses Dirac cones for quasi-momenta
$(k_x,k_y)=(n,m)$. The effective number of Dirac cones is four (easily seen by
shifting the Brillouin zone by $\frac{1}{2}$ in both directions), compared
to two Dirac cones encountered in honeycomb lattices, like in graphene. We notice
that recently it was shown how more complex lattice potentials may render single
Dirac cones within the first Brillouin zone~\cite{DC}.

In the adiabatic regime, a non-zero extrinsic force $F_{ex} {\bf e}_x$ is
treated by using the same approximation as above, yielding
\begin{equation}
u_{\nu,k_x,k_y}(x,y)\rightarrow u_{\nu,k_{x}-F_{ex}t,k_y}(x,y).
\end{equation}
In the presence of a periodic potential, this adiabatic
assumption has become known as the {\it acceleration
theorem}~\cite{acc}. It implies that no population is transferred
between the individual bands $\nu$. As a direct consequence, an initial
state with well localized quasi-momentum within one single band will show
an oscillatory motion rather than a constantly accelerating one, once the
extrinsic force $F_{ex} {\bf e}_x$ has been turned on. These are the well studied
Bloch oscillations~\cite{bloch}, which have been experimentally verified in a
variety of systems, such as cold atomic gases~\cite{blochexp1},
BECs~\cite{blochexp2}, semiconductor superlattices~\cite{blochexp3},
waveguide arrays~\cite{blochexp4}, and photonic crystals~\cite{blochexp5}.

BOs are clearly an outcome of the periodic potential and its characteristic
energy spectrum possessing gaps of forbidden energies at the center and edges of the
Brillouin zone. As the amplitude of the potential $V$ vanishes, the gaps close
and the acceleration theorem breaks down. In particular, for weak lattice
amplitudes, the gap between the first two energy bands scales as $\sim V$. By
linearizing the dispersions around the band gap and replacing $k_x$ by
$k_{x}-F_{ex}t$, one obtains a realization of the celebrated {\it Landau-Zener
model}~\cite{lz}. Here, $k_{x}$ is the initial quasi-momentum in the $x$-direction.
This model is analytically solvable, giving the probability of population
transfer between the two states of the corresponding bands. It has been shown
that the decay of BOs can be correctly estimated using the Landau-Zener
theory~\cite{blochexp2,lzbreak}.

In the present model, the situation becomes qualitatively different.
At the Dirac points, the gap vanishes and the acceleration theorem
cannot be imposed. However, from Fig.~\ref{fig3} it is seen that the
Dirac cones appear between the first two bands, and not between the
second and third band. Hence, assuming a state initially prepared on
the lowest band one may expect predominant population transfer
between the $\nu=1$ and $\nu=2$ bands and not to higher bands.
Thereby, further acceleration is hindered by the gap  between the second
and third band and spin-dependent BOs may still be encountered.
In other words, a modified acceleration theorem involving pair of bands
could result. In fact, as already mentioned, the acceleration theorem
is a specific example of the more general adiabatic theorem, and in
molecular and chemical physics it has long been known that adiabaticity
breaks down in the vicinity of a conical intersection~\cite{bear}. In
this context, it should be kept in mind that adiabaticity is a widespread
concept in physics, and the general definition concerns systems whose
dynamics is governed by an explicitly time-dependent Hamiltonian.
Adiabaticity, as it appears in the acceleration
theorem, is in this sense more reminiscent of the {\it
Born-Oppenheimer approximation}~\cite{bear}, applicable to time-independent
Hamiltonians describing light and heavy quantal degrees of freedom corresponding
here to the spin and the spatial motion, respectively, of the particles.

\section{Numerical results}\label{sec3}

Our numerical analysis will be restricted to a system consisting of weakly interacting
atoms, for which a
mean-field description is expected to capture the phenomena we are interested
in. In the $s$-wave scattering regime, we thereby consider dynamics rendered
by a spinor Gross-Pitaevskii equation~\cite{jonaserik,bec2}
\begin{widetext}
\begin{equation}\label{GP}
\begin{array}{lll}
i \displaystyle{\frac{\partial}{\partial t}\Psi(x,y,t)} & = &
\Bigg\{-\displaystyle{\left(\frac{\partial^2}{\partial
x^2}+\frac{\partial^2}{\partial y^2}\right)} +
V\left(\cos^2(x)+\cos^2(y)\right) -F_{ex}x+v_{so}\left[\begin{array}{cc} 0 & -i\frac{\partial}{\partial x}
-\frac{\partial}{\partial y}\\
-i\frac{\partial}{\partial x}+\frac{\partial}{\partial y} &
0\end{array}\right]
\\ \\
& & +\left[\begin{array}{cc}
g_{11}n_1(x,y,t)+g_{12}n_2(x,y,t) & 0 \\
0 & g_{12}n_1(x,y,t)+g_{22}n_2(x,y,t)
\end{array}\right]
\Bigg\}\Psi(x,y,t).
\end{array}
\end{equation}
\end{widetext}
Here,
\begin{equation}
\Psi(x,y,t) =\left[\begin{array}{c}\langle x,y|
\langle u_1|\Psi(t)\rangle\\
\langle x,y|\langle u_2|\Psi(t)\rangle\end{array}\right]=
\left[\begin{array}{c}\psi_1(x,y,t)\\
\psi_2(x,y,t)\end{array}\right]
\end{equation}
is a two-component spinor wave function of the dark states, $n_i(x,y,t)=|\psi_i(x,y,t)|^2$,
$F_{ex}$ the external longitudinal force, and $g_{11}$, $g_{22}$ and $g_{12}$ are the effective 
$s$-wave scattering amplitudes related to the
scattering lengths $a_{i}$ as $g_{i}=N4\pi\hbar^2a_{i}/(3E_R\sqrt{2\pi}l_zm)$,
where $l_z$ is the width of the single site wave function in the $z$-direction.
This definition of $g_{i}$ implies that the wave function has been normalized
as $\int|\Psi(x,y,t)|^2\,dxdy=1$. In the basis of dark states, the scattering
lengths are effective ones different from the regular hyperfine scattering
lengths. Nevertheless, the ratio between the two scattering amplitudes should
approximately be the same regardless of basis, and, in particular, for a spinor
condensate such as that of $^{87}\mathrm{Rb}$ one has 
$g_{11}\approx g_{22}\approx g_{12}$~\cite{scat}.
We will restrict the analysis to repulsive interactions, $g_{12},\,g_{22},\,g_{12}>0$, and we will
furthermore assume $g_{11}=g_{22}=g_{12}\equiv g$ \cite{scat}.

In the absence of a lattice, the SO coupling gives rise to a non-Abelian
Mead-Berry curvature \cite{mead,berry,wz}, which results in a transverse splitting
of spin components, the intrinsic SHE. However, in Figs.~\ref{fig1} and
\ref{fig2} the motion along the longitudinal $x$-direction was either
constant or monotonously accelerating, while for a weakly tilted lattice,
where BOs dominate the longitudinal motion, it is neither clear how the
transverse spin current nor how the trembling Zitterbewegung motion will be
manifested. For instance, will there be sufficient time for a transverse
spin current to be established within a single Bloch period? In
Ref.~\cite{jonashall}, the intrinsic anomalous Hall effect was studied
in a single particle system confined by an harmonic potential and hence
the motion was as well oscillatory as in the case of BOs. The transverse
current appeared as a rotation on top of the harmonic oscillatory motion,
i.e., in polar coordinates $(r,\varphi)$ the gauge potential causes a non-zero
current in the $\varphi$-direction. Such current would indeed be zero for an
SO coupling possessing an Abelian structure rather than non-Abelian,
e.g. for couplings on the form $v_{so}\left(\hat{\sigma}_\alpha\hat{p}_x +
\hat{\sigma}_\alpha\hat{p}_y\right)$ with $\alpha=x,\,y,\,z$. It is clear
that a corresponding rotational current cannot be strictly encountered in
the present system since exact polar symmetry is broken by the tilting of
the lattice and by the lattice in itself.

The time evolution of the system is solved by employing the
split-operator approach~\cite{split}. The method relies on
factorizing the time-evolution operator into a spatial part and a
momentum part valid in the limit of infinitely small time-step
propagation. One convenient feature of this method is that the
wave function is in principle obtained at any instant of time in
both position $\Psi(x,y,t)$ and momentum $\tilde{\Psi}(p_x,p_y,t)$ space.
Furthermore, for a system bounded from below, the approach is capable of
giving an approximate ground state of the system by simply propagating
it sufficiently long in imaginary time.

\begin{figure}[ht]
\begin{center}
\includegraphics[width=8cm]{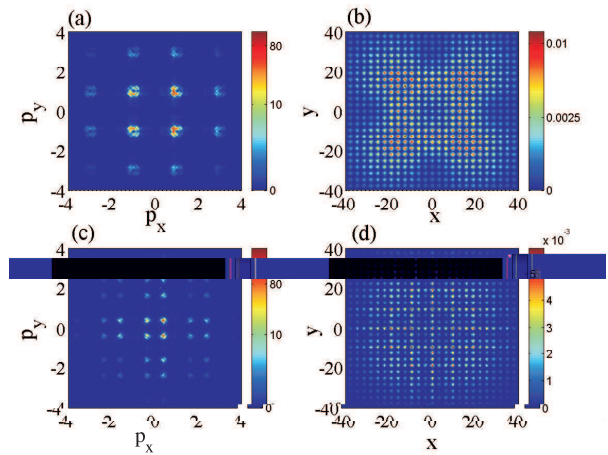}
\caption{(Color online) Absolute value of the ground state
wave function in momentum space (a) and (c), and in position
space (b) and (d). The dimensionless parameters are $V=5$, $F_{ex} = 0$,
and $g=1$ for all plots, $v_{so}=2$ in the upper plots (a) and (b), and
$v_{so}=1$ in the lower plots (c) and (d). Thus, the lower plots
correspond to the dispersions depicted in Fig.~\ref{fig3}. }
\label{fig5}
\end{center}
\end{figure}

Our numerical considerations will focus on propagation of an initial
(approximated) ground state for the untilted ($F_{ex}=0$) lattice. That is, we consider
the ground state of a spin-orbit coupled BEC in an optical lattice, and
then we introduce a quantum quench by suddenly switch on a linear force
in the $x$-direction. Hence, we begin by propagating an ansatz function
in imaginary time with zero extrinsic force $F_{ex}=0$. When convergence
of the imaginary time evolution has been reached, we switch to real time
propagation, including the extrinsic force by adding the potential
term $-F_{ex}\hat{x}$ to the Hamiltonian. As predicted by the
acceleration theorem, for short times the weak forces $F_{ex}$ will induce
a linear increase in the quasi-momentum $k_x$. In contrast to BOs in a
regular square lattice, here the SO coupling implies that also the quasi-momentum
$k_y$ is indirectly affected by the force $F_{ex}$. At the same time, the
build up of motion in the $y$-direction induces an intrinsic force $F_{SO}$
along the longitudinal $x$-direction of BOs. In addition, the linear dispersions
induce a trembling Zitterbewegung motion. It is thereby likely to expect a
complex motion arising from the interplay between the two forces
related to SHE and BOs and the outcomes from linear dispersions.

Figure \ref{fig5} displays the absolute values of the initial ground state
wave function achieved via imaginary time evolution for $F_{ex}=0$. The
momentum wave functions depicted in (a) and (c) demonstrate the Jahn-Teller
effect as the ground state populates non-zero momentum states~\cite{jt}.
The lower two plots (c) and (d) correspond to the parameters of the two
lowest Bloch bands shown in Fig.~\ref{fig3}. The fact that the ground state
obtained from this numerical method is not symmetric, as seen from the plots,
is believed to have a two-fold cause. First, it seems to derive from the fact
that the imaginary time evolution is taken only up to a point where the state
evolution has approximately converged, and, secondly, from the fact that
even small fluctuations can result in an asymmetric ground state, due the
four-fold degeneracy of the ground state of our Hamiltonian. In fact, the complete
convergence of the ground state has not been reached since the position
space wave functions depicted in (b) and (d) are localized in contradiction
to the true ground state having a Bloch structure, and furthermore since the
momentum wave functions have a finite width. This, however, should not be seen
as a shortcoming since normally one would cool the atomic gas in the presence
of a confining potential which would result in a localized gas. In addition,
if the propagated state extends over the full lattice one would run into
troubles when quantities such as the mean spatial position is calculated
numerically. Note that the momentum wave functions of (a) and (c) are different
from the quasi-momentum as is evident in the multiple Bragg components separated
by $\Delta p_x,\,\Delta p_y=\pm1$. In other words, even though the momentum
wave functions extend beyond the first Brillouin zone, it populates, to a
very good approximation, only the lowest band. We also note that the position
space ground state wave functions, shown in (b) and (d), predominantly populate
the individual lattice sites. Hence, only judging from Figs.~\ref{fig5} (b)
and (d) it would be difficult to predict the presence of a SO coupling and
the Dirac cones that derive from it, which instead manifest themselves in
momentum space in terms of a Jahn-Teller effect.

\begin{figure}[ht]
\begin{center}
\includegraphics[width=8cm]{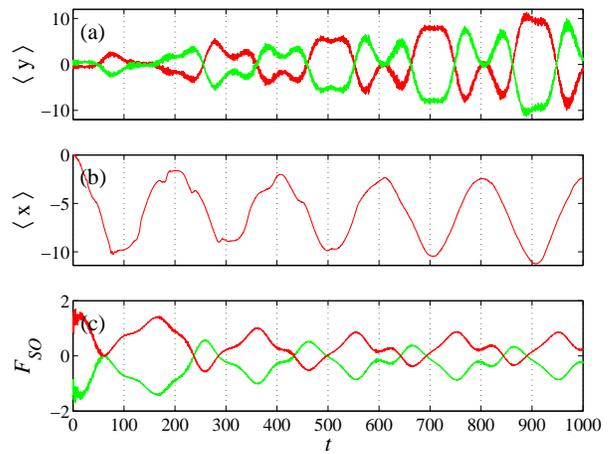}
\caption{(Color online) Average transverse position $\langle\hat{y}\rangle$
(a), average longitudinal position $\langle\hat{x}\rangle$ (b), and
average intrinsic force $F_{SO}$ in the transverse direction (c). Red lines
correspond to the dark state $|u_1\rangle$ while green lines to $|u_2\rangle$.
For $\langle\hat{x}\rangle$ the two lines are identical because of symmetry reasons.
The dimensionless parameters are the same as in Fig.~\ref{fig5} (a) and (b), but
now with a non-zero extrinsic force, $F_{ex}=0.01$. } \label{fig6}
\end{center}
\end{figure}

In order to analyze the interplay between the transverse and
longitudinal forces, ${\bf F}_{SO}$ and $F_{ex} {\bf e}_x$, we calculate
\begin{equation}\label{aver}
\begin{array}{l}
\displaystyle{\langle\hat{y}\rangle_i=\frac{1}{N_i}\int|\psi_i(x,y,t)|^2y\,dxdy}, \\ \\
\displaystyle{\langle\hat{x}\rangle_i=\frac{1}{N_i}\int|\psi_i(x,y,t)|^2x\,dxdy}, \\ \\
\displaystyle{\langle F_{SO}^{(y)}\rangle_i = -g_{so}^2\frac{1}{N_i}\int|\tilde{\psi}_i
(p_x,p_y,t)|^2p_x\,dp_xdp_y},
\end{array}
\end{equation}
with $i=1,\,2$ corresponding to averages in terms of the two dark states $|u_1\rangle$
and $|u_2\rangle$, respectively, $N_j = \iint |\psi_i (x,y,t)|^2 dx dy$, and the last
quantity is the intrinsic force in the $y$-direction. We find that
$\langle\hat{x}\rangle_1=\langle\hat{x}\rangle_2$ (up to numerical accuracy) and
therefore plot only one of the two components.

\begin{figure}[ht]
\begin{center}
\includegraphics[width=8cm]{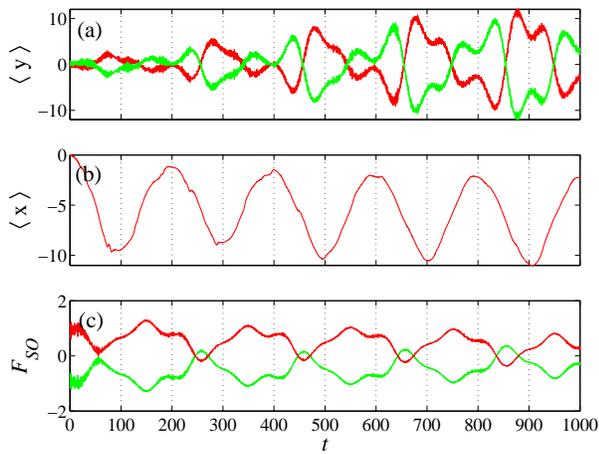}
\caption{(Color online) Same as in Fig.~\ref{fig6}, but with an increased atom-atom
interaction; $g=10$. } \label{fig7}
\end{center}
\end{figure}

Our numerical results are presented in Figs.~\ref{fig6} and
\ref{fig7}. The difference between the two figures lies in the scattering
amplitude $g$. In Fig.~\ref{fig7} the effective scattering amplitude
$g$ is already as large as 10 recoil energies $E_R$ and we can thereby conclude that
the spin current is rather insensitive to the value of $g$. It is generally
known that strong non-linearity can qualitatively change system properties.
In BECs it leads to phenomena such as soliton and vortex creation. Normally
the greatest impact of the non-linearity is seen around curve crossings~\cite{NLcc}.
Beyond a critical value of the non-linearity (in our case $g>g_{crit}$),
in one dimension swallowtail loops are formed in the vicinity of the curve crossings characterizing
additional stationary non-linear states, e.g. solitons. Up to this critical value
the dynamics seems rather insensitive at least on a mean-field level. However,
formation of these loops beyond the critical value implies breakdown of the
acceleration theorem~\cite{atbreak}. To date, it seems that analysis of these
effects have been restricted to one dimensional situations and it is therefore
not fully understood how non-linearity affects the structure of the Dirac cones
appearing in two dimensions. Nonetheless, we have numerically found that the
general structure of Figs.~\ref{fig6} and \ref{fig7} drastically change for
very large values of $g$. All present BO experiments, however, are performed
in the weakly interacting regime where it is known that loops are absent, and
we thereby restrict our analysis to these experimentally more relevant parameter ranges.

We see from both Fig.~\ref{fig6} and \ref{fig7} that the $x$ and $y$ motions
are indeed correlated; when the Bloch oscillation reaches its turning points,
the spin separation in the $y$-direction is roughly maximal and as the velocity
in the longitudinal $x$-direction changes sign, the spin separation decreases
due to the reversed sign of the intrinsic force. When the longitudinal
velocity is the greatest (equivalent to maxima of the intrinsic force
$F_{SO}$), the spin current is reversed. More precisely, as the atoms begin
to accelerate in the tilted lattice the quasi-momentum $k_x$ increases linearly
in time and a spin current is built up in the transverse direction. However,
after half a Bloch period the quasi-momentum reaches the Brillouin boundary
and the $x$-motion is stopped and then reversed. Consequently, also the spin
current is reversed and one encounters a decreasing spin current in the
$y$-direction. Finally, the spin current has dropped to zero and builds up
in the opposite direction instead. Hence, the spin current follows the same
oscillatory behavior as the BOs.

One may argue that a large spin separation should be
hindered by the alternating sign of the velocity in the $x$-direction.
Nonetheless, from our numerical results depicted in Figs.~\ref{fig6} and
\ref{fig7}, we note that the spin separation in the $y$-direction is relatively
large. Indeed, $\langle \hat{y}\rangle$ exceeds several site-spacings and
furthermore this separation amplitude increases in time. The fact that neither
the oscillatory spin nor the oscillatory longitudinal current show a clean
perfect harmonic pattern derives from, at the one hand possible non-adiabatic
effects and the breakdown of the acceleration theorem, and at the other that
the transverse current induces an intrinsic force $F_{SO}^{(x)}$ in the
longitudinal direction which adds to the extrinsic force $F_{ex} {\bf e}_x$.
The full coupled dynamics is therefore rather complex, but nonetheless, our
figures make clear the overall in-phase oscillations between the two currents.

On top of the oscillations of $\langle\hat{x}\rangle_i$, there is a slight
drift toward negative values of $x$. This is an outcome of non-adiabaticity
and breakdown of the acceleration theorem. However, this drift is rather
small and the dynamics is therefore predominantly adiabatic. This drift can
be suppressed by increasing the lattice amplitude $V$ on the cost that the
amplitude of longitudinal oscillations decreases. It would not, however,
imply a diminishing SHE since it is related to the longitudinal momentum and
not position.

A closer look at the two figures reveals that in the transverse motions there
is an additional oscillating motion different from the spin current being in
phase with the BOs. This is believed to be Zitterbewegung clearly seen in the
freely evolving wavepackets of Fig.~\ref{fig1}. Contrary to that figure, here
these oscillations do not collapse suggesting that it might be preferable to
detect the Zitterbewegung effect within a tilted lattice, compared to the
case of freely propagating atoms. The survival of Zitterbewegung should
as well derive from the alternating motion due to the optical lattice.

Figures \ref{fig6} and \ref{fig7} have been calculated for an approximated
initial ground state of the untilted lattice. As already pointed out,
experimentally the condensate might be prepared in the presence of an
external trap, and then the optical lattice as well as the coupling of
the internal states are turned on in such a way that the condensate is
not in a true ground state of the coupled system. To analyze the effect
of such excitations, we have redone the simulations of Figs.~\ref{fig6}
and \ref{fig7} for an initial Gaussian momentum state of a form depicted
in Eq.~(\ref{gaussin}) with $p_0=0$ and $\Delta_p=0.1$. The results are very
similar to the ones of Figs.~\ref{fig6} and \ref{fig7}, and we therefore
do not present them here. We have also investigated the effect of a shallow
harmonic potential in the transverse direction during the evolution in the
tilted lattice. Again, a transverse spin current is obtained, however this
time the amplitude of transverse oscillations is decreased due to the confining
potential. This reduction of spin separation would indeed be present in any
experiment utilizing a confining trapping potential. The idea of employing an
optical lattice potential as a trapping potential circumvent such shorting.
In order to achieve trapping in this case one must stay within the BO regime,
i.e. adiabatic. As Figs.~\ref{fig6} and \ref{fig7} demonstrates, even though
the atoms oscillate a relatively large spin current can build up during time.

In relation to the measurability of the spin separation, we may note that even
though $|\langle\hat{y}\rangle_1-\langle\hat{y}\rangle_2|$ is considerably
larger than the lattice spacing, it does not imply that the two wave functions
$\psi_1(x,y,t)$ and $\psi_2(x,y,t)$ do not spatially overlap. In fact, the
overlap is relatively large and fluorescent measurement of a single experimental
run would therefore not contain sufficient information to reveal the spin current.
However, this is hardly never the case in true experimental realizations and
instead an ensemble of measurements should be considered, where the ensemble
average gives $|\langle\hat{y}\rangle_1-\langle\hat{y}\rangle_2|$. Since this
value is presumably larger than the lattice spacing, it should indeed be measurable~\cite{sitemeas}.
In the cold atom community, time-of-flight measurements are often employed, in
which it is the atomic momentum density that is being measured. If the separation
of the momentum wave functions is non-zero, the position space wave functions
$\psi_1(x,y,t)$ and $\psi_2(x,y,t)$ will set apart further during the time-of-flight.
If this separation is faster than the intrinsic wave function broadening, a
finite time-of-flight measurement may enhance the detection efficiency.

\section{Conclusion}\label{sec4}
In this paper, we have considered an ultracold atomic system possessing three
well-studied properties of condensed matter theory, Bloch oscillations, intrinsic
spin Hall currents, and Zitterbewegung.Our motivation for this study comes from 
the fact that Bloch oscillations have been verified in the
same type of setups \cite{blochexp2,blochexp1}, light induced gauge potentials
have been demonstrated \cite{lightexp1,lightexp2,nonab}, and Zitterbewegung have been
detected in ion-trap experiments~\cite{solano}.

The spin-Hall effect (SHE) and the Bloch oscillations (BO) in our system are 
driven by two perpendicular forces: the extrinsic force appearing due to tilting 
of the lattice potential and the intrinsic force
originating from the effective gauge potential caused by the spin-orbit (SO)
coupling. We demonstrated that the interplay between these two forces gives
rise to a correlated motion in the $xy$-plane, and despite the fact that the
BOs are expected to prevent a build up of large transverse spin currents, we
argued that the SHE should indeed be detectable. We utilized system parameters
that should be achievable in current experiments, e.g., the amplitude of the
lattice potential is five recoil energies compared to the experimentally utilized
ones of Ref.~\cite{blochexp2} ranging from one to 14 recoil energies, and the SO
coupling two recoil energies compared to 8 recoil energies in Ref.~\cite{lightexp2}.
For a potential amplitude as large as $14E_r$, the dispersions are almost flat and
longitudinal motion is almost frozen. The momentum along the longitudinal direction
do, however, oscillate within the Brillouin size and, interestingly and somewhat
surprisingly, one would therefore see an oscillating transverse spin Hall current 
despite lack of longitudinal currents. We furthermore showed that the effects are 
insensitive to experimentally
relevant atom-atom interaction strengths. Entering a strongly interacting gas
seemed to affect our results and we argue that this is likely to derive
from a non-linearity induced breakdown of adiabaticity around the Dirac points. An
additional interesting aspect being discussed is the analogue of relativistic
Zitterbewegung. Our numerical results indicate that it may be preferable to
consider an optical lattice configuration when studying this effect since it
persists over longer time periods. The optical lattice convey as well the
possibility to avoid confining trapping potentials preventing large spin currents.

We should also point out that the system presented in this paper allows
for many extensions and different phenomena to be studied. For
example, vortex generation in lattices~\cite{nonab,vortex}, Bose-Einstein condensation
into non-zero momentum states~\cite{wu}, topologically ordered states of
matter~\cite{top}, or more strongly correlated systems, by means of second
quantization of the mean-field Hamiltonian, may find realization in atoms
possessing an internal tripod structure and trapped in optical lattices. The
flexibility permits as well for various extensions, for example time-dependent
lattices or SO couplings. A recent paper considered Zitterbewegung of ultracold
atoms for a driven SO coupling~\cite{OH}. The same idea can be utilized in
order to increase the transverse spin current in our model.

\begin{acknowledgments}
JL acknowledges support from the MEC program (FIS2005-04627). JL and ES acknowledge 
financial support from the Swedish Research Council.
\end{acknowledgments}


\begin{thebibliography}{999}
\bibitem{trap} It is worth pointing out that even if a confining trap is absent in 
the actual BO experiments, it is present during the condensation process and as it 
is turned off before releasing the condensate in the tilted lattice it implies, in 
the local density approximation, that the chemical potential varies within the extent 
of the atomic cloud which renders a source for decoherence, see~\cite{blochexp2}.
\end{thebibliography}
\end{document}